\documentclass[a4paper,11pt]{article}
\pdfoutput=1
\usepackage{jcappub} 
\allowdisplaybreaks

\newcommand{\Planck}{{\slshape Planck~}}

\newcommand{\bk}{\mathbf{k}}
\newcommand{\bq}{\mathbf{q}}
\newcommand{\br}{\mathbf{r}}
\newcommand{\bs}{\mathbf{s}}
\newcommand{\bt}{\mathbf{t}}
\newcommand{\bu}{\mathbf{u}}

\newcommand{\Pl}{P}

\begin{document}

\title{The two and three-loop matter bispectrum in perturbation theories}
\author[a]{Andrei Lazanu,}
\author[a,b,c]{Michele Liguori}

\affiliation[a]{INFN, Sezione di Padova, via Marzolo 8, I-35131, Padova, Italy}
\affiliation[b]{Dipartimento di Fisica e Astronomia ``G. Galilei'',Universit\`a degli Studi di Padova, via Marzolo 8, I-35131, Padova, Italy}
\affiliation[c]{INAF-Osservatorio Astronomico di Padova, vicolo dell Osservatorio 5, I-35122, Padova, Italy}

\emailAdd{Andrei.Lazanu@pd.infn.it}

\abstract{We evaluate for the first time the dark matter bispectrum of large-scale structure at two loops in the Standard Perturbation Theory and at three loops in the Renormalised Perturbation Theory (\textsc{MPTbreeze} formalism), removing in each case the leading divergences in the integrals in order to make them infrared-safe. We show that  the Standard Perturbation Theory at two loops can be employed to model the matter bispectrum further into the quasi-nonlinear regime compared to one loop, up to $k_{\text{max}} \sim 0.1 \, h/{\text{Mpc}}$ at $z = 0$, but without reaching a high level of accuracy. In the case of the \textsc{MPTbreeze} method, we show that its bispectra decay at smaller and smaller scales with increasing loop order, but with smaller improvements.  At three loops, this model predicts the bispectrum accurately up to scales $k_{\text{max}} \sim 0.17 \, h/{\text{Mpc}}$ at $z = 0$ and $k_{\text{max}} \sim 0.24 \, h/{\text{Mpc}}$ at $z = 1$.} 

\maketitle
\flushbottom

\section{Introduction}

In recent years, cosmological observations have achieved an unprecedented level of accuracy. Probes measuring the Cosmic Microwave Background (CMB) such as \Planck \cite{Planck2015} have confirmed the inflationary paradigm and have been able to measure cosmological parameters with exquisite precision, showing that the six-parameter $\Lambda$CDM model provides an accurate description of the data. Nevertheless, deviations from this model as well as additional ingredients in the Universe are still allowed, but their parameters are constrained by cosmological measurements. Most of these results have been derived using statistical tools involving two-point correlation functions (power spectra). Higher-order statistics, such as the three-point correlation function -- bispectrum, can be used to extract additional and complementary information. Compared to the power spectrum, bispectra are much more difficult to extract and to analyse due to their three-dimensional nature, but this is also an important advantage, as one can also use the shape information encoded.  With \textit{Planck}, the CMB has been exploited almost to the cosmic variance limit, and the late-time dark matter distribution -- large scale structure (LSS) of the Universe has been increasingly used to extract information from cosmological probes. Apart from the challenge of finding new physics, it can be used as an independent confirmation of the validity of existing theories. Compared to the CMB, it contains significantly more information, since the CMB only provides data from the surface of last scattering, while LSS also encodes time (redshift) information. This information is however much more challenging to extract due to a number of reasons: the nonlinear nature of structure formation, the relationship between observed quantities -- galaxies and the underlying dark matter distribution -- the bias, redshift space effects.

Accurate modelling of all these effects is now a priority since data from a large number of galaxy probes will be available in the near future, such as DESI \cite{DESI}, Euclid \cite{Euclid}, LSST \cite{LSST}. Similarly to the CMB, most of the work and analyses have been devoted to the study of power spectrum statistics. The bispectrum can be used in LSS to disentangle degeneracies between bias parameters and primordial non-Gaussianity and to place stringent constraints on various non-Gaussian shapes \cite{Scoccimarro2001,Gilmarin2017,Tellarini2016,Karagiannis2018}.

Even in the absence of primordial non-Gaussianity, the matter bispectrum is non-zero due to the nonlinear structure growth. Therefore, the study of such higher-order statistics is important not only for investigating primordial non-Gaussianity, but also more generally in LSS. This, together with an accurate bias model, can be used to compute the bispectrum of galaxies and to compare with observations, or it can be used directly with weak lensing surveys.

The complex, nonlinear, three-dimensional nature of the dark matter bispectrum statistics makes it difficult to model, with the linear prediction for dark matter statistics being valid on large scales. To go beyond such scales, one can rely on perturbation theory techniques, that describe clustering on quasi-nonlinear scales, use halo models \cite{Cooray2002} or other phenomenological models \cite{Lazanu2016,Lazanu2017a}, or use $N$-body simulations. However, bispectrum simulations are expensive to run and checks with analytical techniques are desirable. 

Perturbation theories generally consider small perturbations around the homogeneous background of the Universe \cite{Bernardeau2002}, and have been successfully used for the matter power spectra up to three loops \cite{Blas2004}, but for bispectra these have been mostly confined to one loop, with a numerical implementation of a two-loop bispectrum in renormalised perturbation theory \cite{Lazanu2016,Lazanu2017}. 

The numerical computation of higher-loop polyspectra is numerically challenging due to the large number of terms, with high amplitudes, but alternating signs, the high-order perturbation theory kernels, as well as to the fact that these kernels have divergences that cancel out between various terms, but that must be removed prior to integration in order to ensure that an accurate result is obtained. This technique has been applied to the power spectrum \cite{Carrasco2014} and bispectrum \cite{Angulo2015,Lazanu2016}.

The Eulerian Standard Perturbation Theory (SPT) \cite{Jain1994, Bernardeau2002}, represents the most basic extension of the linear theory, but in order to improve the accuracy of the modelling, one must use renormalised techniques, such as the renormalised perturbation theory and its simplification \textsc{MPTbreeze} \cite{Scoccimarro2001b,Crocce2006,Crocce2006b,Crocce2008,Bernardeau2008,Crocce2012,Bernardeau2012}, the Resummed Lagrangian Perturbation Theory \cite{Okamura2011,Rampf2012,Rampf2012b} or add counterterms through the Effective Field Theory of LSS \cite{Carrasco2012,Pajer2013,Carrasco2014,Porto2014,Senatore2015}.

In this work, we concentrate on the modelling of the matter bispectrum of large scale structure. We investigate the bispectrum predictions of perturbative methods up to two loops in SPT and three loops in \textsc{MPTbreeze}. In the next sections we describe the formalism to determine the matter bispectrum in the two theories, for Gaussian initial conditions, by showing how to  remove the divergences in the integrands appearing in the kernel functions and we compute the bispectra numerically, showing that these models can be useful in extending the range of validity of perturbation theories.

\section{Standard Perturbation Theory}

To describe matter clustering we start by defining the matter overdensity $\delta \equiv (\rho-\bar{\rho})/\bar{\rho}$ in Fourier space, where $\rho$ is the matter density and $\bar{\rho}$ is its spatial average. The power spectra and bispectra are then given by
\begin{align}
\label{ps}
\langle	\delta (\textbf{k}_1) \delta (\textbf{k}_2) \rangle &= (2\pi)^3 \delta_D (\textbf{k}_1 + \textbf{k}_2) P(k) \, , \\
\langle	\delta (\textbf{k}_1) \delta (\textbf{k}_2) \delta (\textbf{k}_3) \rangle &= (2 \pi)^3 \delta_D (\textbf{k}_1 + \textbf{k}_2 + \textbf{k}_3) B(k_1,k_2,k_3) \, ,
\label{bis}
\end{align}
where $\textbf{k}_i$ are wavevectors with magnitudes $k_i$ and $\delta_D$ is the Dirac delta function. In SPT one considers a fluid description of dark matter \cite{Bernardeau2002} to expand the matter overdensity  around the homogeneous background. Using the Euler equation, perturbative solutions for $\delta$ and $\theta$ (the divergence of the velocity field) are derived, under the assumptions  $|\delta| \ll 1$ and $|\theta| \ll 1$. In a $\Lambda$CDM universe, they can be expanded as
\begin{align}
\label{exp1}
{\delta} \left(\textbf{k},\tau\right)&=\sum_{n=1}^{\infty}D^n\left(z\right)\delta_n\left(\textbf{k}\right) \, ,\\
{\theta} \left(\textbf{k},\tau\right)&=-\mathcal{H}\sum_{n=1}^{\infty}D^n\left(z\right)\theta_n\left(\textbf{k}\right) \, ,
\label{exp2}
\end{align}
where $D\left(z\right)$ is the linear growth factor normalised to one today and $\mathcal{H}$ is the conformal Hubble rate. The solutions for $\delta_n$ and $\theta_n$ are given by
\begin{align}
\label{deltanf}
\delta_n\left(\textbf{k}\right)=&\int d^3\textbf{q}_1 \ldots \int d^3\textbf{q}_nF_n^{(s)}\left(\textbf{q}_1, \ldots ,\textbf{q}_n\right)  \delta_1\left(\textbf{q}_1\right) \cdots \delta_1\left(\textbf{q}_n\right) \delta_D(\textbf{k}-\textbf{q}_1-\ldots -\textbf{q}_n) \, , \\
\theta_n\left(\textbf{k}\right)=&\int d^3\textbf{q}_1 \ldots \int d^3\textbf{q}_nG_n^{(s)}\left(\textbf{q}_1, \ldots ,\textbf{q}_n\right)  \delta_1\left(\textbf{q}_1\right) \cdots \delta_1\left(\textbf{q}_n\right)\delta_D(\textbf{k}-\textbf{q}_1-\ldots - \textbf{q}_n) \, ,
\end{align}
with kernels $F_n$ and $G_n$ satisfying the recurrence relations
\begin{align}
\label{Fn}
F_n\left(\textbf{q}_1,\ldots,\textbf{q}_n\right)=\sum_{m=1}^{n-1} &\frac{G_m\left(\textbf{q}_1, \ldots, \textbf{q}_m\right)}{\left(2n+3\right)\left(n-1\right)} 
\left[\left(2n+1\right)\alpha\left(\textbf{k}_1,\textbf{k}_2\right)F_{n-m}\left(\textbf{q}_{m+1},\ldots,\textbf{q}_n\right) \right. \nonumber \\
&+\left.2\beta\left(\textbf{k}_1,\textbf{k}_2\right)G_{n-m}\left(\textbf{q}_{m+1},\ldots,\textbf{q}_n\right) \right] \, , \\
G_n\left(\textbf{q}_1,\ldots,\textbf{q}_n\right)=\sum_{m=1}^{n-1} &\frac{G_m\left(\textbf{q}_1, \ldots, \textbf{q}_m\right)}{\left(2n+3\right)\left(n-1\right)}  
\left[3\alpha\left(\textbf{k}_1,\textbf{k}_2\right)F_{n-m}\left(\textbf{q}_{m+1},\ldots,\textbf{q}_n\right) \right. \nonumber \\
&+\left.2n\beta\left(\textbf{k}_1,\textbf{k}_2\right)G_{n-m}\left(\textbf{q}_{m+1},\ldots,\textbf{q}_n\right) \right] \, ,
\label{Gn}
\end{align}
with $F_1 = G_1 = 1$, $\textbf{k}_1=\textbf{q}_1+\ldots+\textbf{q}_m$, and $\textbf{k}_2=\textbf{q}_{m+1}+\ldots+\textbf{q}_n$ and
\begin{align}
\label{alpha}
\alpha\left(\bk_1,\bk_2\right)&=\frac{(\bk_1+\bk_2) \cdot \bk_{1}}{k_1^2} \, , \\
\beta\left(\bk_1,\bk_2\right)&= \frac{|\bk_1+\bk_2| \left(\bk_{1} \cdot \bk_{2} \right)}{2k_1^2k_2^2} \, .
\label{beta}
\end{align}

\noindent
The symmetrised versions of these kernels are
\begin{align}
\label{Fns}
F_n^{(s)}\left(\textbf{q}_1,\ldots,\textbf{q}_n\right)&=\frac{1}{n!}\sum_{\text{all perms}} F_n\left(\textbf{q}_{i_1},\ldots,\textbf{q}_{i_n}\right)  \, ,\\
G_n^{(s)}\left(\textbf{q}_1,\ldots,\textbf{q}_n\right)&=\frac{1}{n!}\sum_{\text{all perms}} G_n\left(\textbf{q}_{i_1},\ldots,\textbf{q}_{i_n}\right) \, .
\end{align}
To derive correlation functions for the matter overdensity, we plug in the expansion (\ref{exp1}) into the relevant correlator, 
\begin{align}
 \langle \delta(\bk_1) \delta(\bk_2) \delta(\bk_3)\rangle 
 &=\langle (\delta_1(\bk_1)+\delta_2(\bk_1)+ \ldots) (\delta_1(\bk_1)+\delta_2(\bk_2)+ \ldots) (\delta_1(\bk_3)+\delta_2(\bk_3)+ \ldots)\rangle \, .
\end{align}
After expanding the brackets and using Eq. (\ref{deltanf}), we use Wick's theorem for Gaussian fields for $\delta_1$, which states that the $n$-point correlation function can be split into a sum of products of two-point  (linear) correlation functions for even $n$, while for odd $n$ the correlation function is 0.

In analogy to quantum field theory, the order of the expansion can be interpreted as a `loop expansion'. The bispectrum expansion in SPT is given in terms of the kernels $F_n^{(s)}$ and the linear power spectrum $P(k)$ as
\begin{equation}
 B(k_1,k_2,k_3)=(B_{\text{tree}}+B_{\text{1 loop}}+B_{\text{2 loops}} + \ldots )(k_1,k_2,k_3) \, .
\end{equation}
For convenience, we present the results at redshift $z=0$, noting that each linear power spectrum $\Pl$ carries a factor of $D^2(z)$, and hence the tree-level bispectrum will be multiplied by $D^4(z)$, the one-loop bispectrum by  $D^6(z)$ and the two-loop bispectrum by $D^8(z)$. The power spectrum is evaluated at $z=0$. The tree-level bispectrum is
\begin{align}
 B_{\text{tree}}(k_1,k_2,k_3)= 2F_2^{(s)}\left(\textbf{k}_1,\textbf{k}_2\right)\Pl(k_1) \Pl(k_2) + \text{2 perms.} \, .
\end{align}
The one-loop bispectrum contains four terms, given by
\begin{equation}
B_{\text{1-loop}}^{\mathrm{SPT}} =B_{222}+B_{321}^{(I)}+B_{321}^{(II)}+B_{411} \, ,
\end{equation}
which have the following expressions:
\begin{align}
\label{b222}
B_{222}\left(k_1,k_2,k_3, z\right)&=8 \int_{\textbf{q}}P_{\text{lin}}\left(q\right)P_{\text{lin}}\left(|\textbf{k}_2-\textbf{q}|\right)  
P_{\text{lin}}\left(|\textbf{k}_3+\textbf{q}|\right) \nonumber \\
&\times F_2^{\left(s\right)}\left(-\textbf{q},\textbf{k}_3+\textbf{q}\right)F_2^{\left(s\right)}\left(\textbf{k}_3+\textbf{q},\textbf{k}_2-\textbf{q}\right)F_2^{\left(s\right)}\left(\textbf{k}_2-\textbf{q},\textbf{q}\right)  \, ,\\
\label{b321i}
B_{321}^{(I)}\left(k_1,k_2,k_3, z\right) &= 6P_{\text{lin}}\left(k_3\right)\int_{\textbf{q}}P_{\text{lin}}\left(|\textbf{k}_2-\textbf{q}|\right)P_{\text{lin}}\left(q\right)  \nonumber \\
&\times F_3^{\left(s\right)}\left(-\textbf{q},-\textbf{k}_2+\textbf{q},-\textbf{k}_3\right)F_2^{\left(s\right)}\left(\textbf{k}_2-\textbf{q},\textbf{q}\right) + \text{5 perms.} \, ,\\
B_{321}^{(II)}(k_1,k_2,k_3, z) &= 6  P_{\text{lin}}(k_2) P_{\text{lin}}(k_3)F_2^{(s)}(\textbf{k}_2,\textbf{k}_3)  \nonumber \\
&\times \int_{\textbf{q}}P_{\text{lin}}\left(q\right)F_3^{\left(s\right)}\left(\textbf{k}_3,\textbf{q},-\textbf{q}\right) + \text{5 perms.} \, ,\\
\label{b411}
B_{411}\left(k_1,k_2,k_3, z\right)&=12 P_{\text{lin}}\left(k_2\right)P_{\text{lin}}\left(k_3\right)  
\int_{\textbf{q}}P_{\text{lin}}\left(q\right)F_4^{\left(s\right)}\left(\textbf{q},-\textbf{q},-\textbf{k}_2,-\textbf{k}_3\right) + \text{2 perms.} \, ,
\end{align}
where $\int_{\textbf{q}} \equiv \int \frac{d^3 \bq}{\left(2\pi\right)^3}$. 

\subsection*{Two-Loop Terms}
The two-loop SPT bispectrum is composed of 11 terms, 
\begin{align}
\label{b2loop}
B_{\text{2-loops}}^{\mathrm{SPT}} =& B_{116}+B_{125}^{(I)}+B_{125}^{(II)}+B_{134}^{(I)}+B_{134}^{(II)} \nonumber \\ +&B_{134}^{(III)}+B_{224}^{(I)}+B_{224}^{(II)}+B_{233}^{(I)}+B_{233}^{(II)}+B_{233}^{(III)} \, .
\end{align}
Their expressions are
\begin{align}
\label{b116}
B_{116}(k_1,k_2,k_3)=&90 \int_{\bq,\br} F_6^{(s)}(\bq,-\bq,\br,-\br,-\bk_1,-\bk_2)\Pl(q)\Pl(r)\Pl(k_1)\Pl(k_2) + 2 \text{ perms.} \, , \\
\label{b125i}
B_{125}^I(k_1,k_2,k_3)=&30 \int_{\bq,\br} F_2^{(s)}(-\bk_1,-\bk_3) F_5^{(s)}(\bk_3,\bq,-\bq,\br,-\br) \nonumber \\
& \qquad \quad \times\Pl(q)\Pl(r)\Pl(k_1)\Pl(k_3) + 5 \text{ perms.} \, , \\
\label{b125ii}
B_{125}^{II}(k_1,k_2,k_3)=&60 \int_{\bq,\br} F_2^{(s)}(\bq,\bk_2-\bq) F_5^{(s)}(-\bk_1,-\bq,\bq-\bk_2,\br,-\br) \nonumber \\
& \qquad \quad \times\Pl(q)\Pl(r)\Pl(k_1)\Pl(|\bk_2-\bq|) + 5 \text{ perms.} \, , \\
\label{b134i}
B_{134}^{I}(k_1,k_2,k_3)=&36 \int_{\bq,\br} F_3^{(s)}(-\bk_1,\bq,-\bk_3-\bq) F_4^{(s)}(-\bq,\bk_3+\bq,\br,-\br) \nonumber \\
& \qquad \quad \times   \Pl(k_1) \Pl(|\bk_3+\bq|)\Pl(q)\Pl(r) + 5 \text{ perms.} \, , \\
\label{b134ii}
B_{134}^{II}(k_1,k_2,k_3)=&24 \int_{\bq,\br} F_3^{(s)}(\bq,\br,\bk_2-\bq-\br) F_4^{(s)}(-\bk_1,-\bq,-\br,\bq+\br-\bk_2) \nonumber \\
& \qquad \quad \times    \Pl(k_1)\Pl(|\bk_2-\bq-\br|)\Pl(q)\Pl(r) + 5 \text{ perms.} \, ,\\
\label{b134iii}
B_{134}^{III}(k_1,k_2,k_3)=&36 \int_{\bq,\br} F_3^{(s)}(\bk_2,\bq,-\bq) F_4^{(s)}(-\bk_1,-\bk_2,\br,-\br) \nonumber \\
& \qquad \quad \times     \Pl(k_1) \Pl(k_2)\Pl(q)\Pl(r) + 5 \text{ perms.} \, ,\\
\label{b224i}
B_{224}^I(k_1,k_2,k_3)=&48 \int_{\bq,\br} F_2^{(s)}(\bq,\bk_1-\bq)F_2^{(s)}(-\bq,\bk_2+\bq) F_4^{(s)}(\bq-\bk_1,-\bk_2-\bq,\br,-\br) \nonumber \\
& \qquad \quad \times    \Pl(|\bk_1-\bq|) \Pl(|\bk_2+\bq|)\Pl(q)\Pl(r) + 2 \text{ perms.} \, , \\
\label{b224ii}
B_{224}^{II}(k_1,k_2,k_3)=&24 \int_{\bq,\br} F_2^{(s)}(\bq,\bk_1-\bq)F_2^{(s)}(\br,\bk_2-\br) F_4^{(s)}(-\bq,\bq-\bk_1,-\br,\br-\bk_2) \nonumber \\
& \qquad \quad \times   \Pl(|\bk_1-\bq|) \Pl(|\bk_2-\br|)\Pl(q)\Pl(r) + 2 \text{ perms.}  \, , \\
\label{b233i}
B_{233}^I(k_1,k_2,k_3)=& 36 \int_{\bq,\br} F_2^{(s)}(\bq,\bk_1-\bq)F_3^{(s)}(-\bq,\bq-\bk_1,-\bk_3) F_3^{(s)}(\bk_3,\br,-\br) \nonumber \\
& \qquad \quad \times    \Pl(k_3) \Pl(|\bk_1-\bq|)\Pl(q)\Pl(r) + 2 \text{ perms.} \, , \\
\label{b233ii}
B_{233}^{II}(k_1,k_2,k_3)=& 36 \int_{\bq,\br} F_2^{(s)}(\bq,\bk_1-\bq)F_3^{(s)}(-\bq,\br,\bk_2+\bq-\br) F_3^{(s)}(\bq-\bk_1,-\br,\br-\bk_2-\bq) \nonumber \\
& \qquad \quad \times    \Pl(|\bk_1-\bq|) \Pl(|\bk_2+\bq-\br|)\Pl(q)\Pl(r) + 2 \text{ perms.}  \, ,\\
\label{b233iii}
B_{233}^{III}(k_1,k_2,k_3)=& 18 \int_{\bq,\br} F_2^{(s)}(-\bk_2,-\bk_3)F_3^{(s)}(\bk_2,\bq,-\bq) F_3^{(s)}(\bk_3,\br,-\br) \nonumber \\
& \qquad \quad \times    \Pl(k_2) \Pl(k_3)\Pl(q)\Pl(r) + 2 \text{ perms.} \, .
\end{align}

This expansion for SPT is evaluated numerically and results are shown is Section \ref{sec:res}. However, this loop expansion is not a well defined expansion in the sense that the magnitude of the terms does not decrease with the expansion order and therefore going to higher orders would not necessarily mean a better agreement with simulations. Moreover, the loop integrals involve integrals over infinite domains, where the basic assumption of perturbation theory, $|\delta| \ll 1$, is no longer valid \cite{Bernardeau2002}.

To cure these issues and still use a perturbative approach, a possible way is to use a renormalised technique, which regroups the SPT terms into a convergent expansion. 

\section{Renormalised Perturbation Theory}

The renormalised perturbation theory has been developed in Refs.~\cite{Scoccimarro2001b, Crocce2006, Crocce2006b,Crocce2008, Bernardeau2008, Crocce2012, Bernardeau2012}, where the standard SPT terms are reorganised into a convergent expansion. By defining the two-component vector,
\begin{equation}
\Psi\left(\textbf{k},\eta\right)=\left(\delta\left(\textbf{k},\eta\right),-\theta\left(\textbf{k},\eta\right)/\mathcal{H}\right) \, ,
\end{equation}
the evolution equations can be expressed as
\begin{equation}
\label{eqpsi1}
\partial_\eta \Psi_a\left(\textbf{k},\eta\right)+\Omega_{ab}\left(\textbf{k},\eta\right)=  
\gamma_{abc}^{(s)}\left(\textbf{k},\textbf{k}_1,\textbf{k}_2\right)\Psi_b\left(\textbf{k}_1,\eta\right)\Psi_c\left(\textbf{k},\eta\right) \, ,
\end{equation}
where
\begin{equation}
\Omega_{ab}=\left(\begin{array}{cc}
                   0 & -1/2\\
                   -3/2 & 1/2
                  \end{array}
\right) \, ,
\end{equation}
and $\gamma_{abc}^{(s)}$ is a symmetrised vertex matrix given in terms of the functions $\alpha$ and $\beta$. The solution to Eq. (\ref{eqpsi1}) is given by 
\begin{equation}
\label{eqpsi}
\Psi_a\left(\textbf{k},\eta\right)=g_{ab}\left(\eta\right)\phi\left(\textbf{k}\right)+\int_0^{\eta}d\eta'g_{ab}\left(\eta-\eta'\right)  
\gamma_{bcd}^{(s)}\left(\textbf{k},\textbf{k}_1,\textbf{k}_2\right)\Psi_c\left(\textbf{k}_1,\eta'\right)\Psi_d\left(\textbf{k}_2,\eta'\right) \, ,
\end{equation}
where $g_{ab}$ is the linear propagator, 
\begin{equation*}
g_{ab}\left(\eta\right) = \begin{cases}
\frac{e^{\eta}}{5}\left(\begin{array}{cc}
                   3 & 2\\
                   3 & 2
                  \end{array}\right)-\frac{e^{-3\eta/2}}{5}\left(\begin{array}{cc}
                   -2 & -2\\
                   3 & -3
                  \end{array}\right) &\text{if $\eta>0$}\\
0 &\text{if  $\eta<0$}
\end{cases} \, .
\end{equation*}
Eq.~(\ref{eqpsi}) yields the SPT solution
\begin{equation}
\Psi_a\left(\textbf{k},\eta\right)=\sum_{n=1}^{\infty}\Psi_a^{(n)}\left(\textbf{k},\eta\right) \, ,
\label{sumpsi}
\end{equation}
where
\begin{equation}
\Psi_a^{(n)}\left(\textbf{k},\eta\right)= \int \delta_D\left(\bk-\bk_1 \cdots -\bk_n\right)\mathcal{F}^{(n)}_{aa_1 \cdots a_n}\left(\textbf{k}_1, \cdots, \textbf{k}_n; \eta\right)  
\phi(\textbf{k}_1) \cdots \phi(\textbf{k}_n) \, .
\end{equation}
This formalism can be generalised to account for nonlinearities, by defining a nonlinear propagator
\begin{equation}
G_{ab}\left(k,\eta\right)\delta_D\left(\textbf{k}-\textbf{k}'\right)=\left\langle\frac{\delta\Psi_a(\textbf{k},\eta)}{\delta \phi_b(\textbf{k}')}\right\rangle \, ,
\end{equation}
which can be expressed as an infinite sum
\begin{equation}
G_{ab}\left(k,\eta\right)=g_{ab}\left(k,\eta\right)+\sum_{n=2}^{\infty}\left\langle\frac{\delta\Psi_a^{(n)}(\textbf{k},\eta)}{\delta \phi_b(\textbf{k}')}\right\rangle \, .
\end{equation}
Nonlinearities also modify the vertex functions $\gamma_{abc}$ to a nonlinear vertex function $\Gamma$,
\begin{multline}
\left\langle\frac{\delta^2\Psi_a(\textbf{k},\eta)}{\delta \phi_e(\textbf{k}_1) \delta \phi_f(\textbf{k}_2)}\right\rangle = 2 \int_0^{\eta}ds\int_0^s ds_1 \int_0^s ds_2 G_{ab}\left(\eta-s\right)  \\
\times \Gamma_{bcd}^{(s)}\left(\textbf{k},s;\textbf{k}_1,s_1;\textbf{k}_2,s_2\right)G_{ce}(s_1)G_{df}(s_2) \, .
\end{multline}

\noindent
In the small-scale limit, the infinite series for the propagator can be resummed to \cite{Crocce2006b}
\begin{equation}
G_{ab}(k,a)=g_{ab}(a)\exp\left(-\frac{k^2\sigma_d^2}{2}\right) \, ,
\end{equation}
where $\sigma_d^2=\frac{(a-1)^2}{3}\int \frac{d^3q}{(2\pi)^3}\frac{P_{\text{lin}}(q)}{q^2}$.

In this theory, all the contributions involved are positive and the resummation of the propagator terms gives a well-defined perturbative expansion in the nonlinear regime, but the expressions are complicated and computationally expensive and high-order loop expansions are required to recover correlation functions even on mildly nonlinear scales. A simplification on this theory has been proposed in Refs.~\cite{Bernardeau2008, Crocce2012},  \textsc{MPTbreeze}, where only the late-time propagator is calculated and hence no time integrations are required. 

The nonlinear propagator can be generalised to an arbitrary number of points; the $(p+1)$-point propagator $\Gamma^{(p)}$ is given by
\begin{equation}
\frac{1}{p!}\left\langle \frac{\delta \Psi_a^p\left(\textbf{k},a\right)}{\delta \phi_{b_1}(\textbf{k}_1) \cdots \delta \phi_{b_p}(\textbf{k}_p)}\right\rangle 
= \delta_D\left(\textbf{k}-\textbf{k}_1 -\ldots -\bk_p\right) \Gamma_{ab_1 \ldots b_p}^{(p)}\left(\textbf{k}_1,\ldots,\textbf{k}_p,z\right) \, .
\end{equation}

If only the growing mode initial conditions are considered, the  solutions reduce to the simple expression
\begin{equation}
\Gamma_\delta^{(n)}\left(\textbf{k}_1, \cdots, \textbf{k}_n;z\right) =  \\
D^n\left(z\right)F_n^{(s)}\left(\textbf{k}_1, \cdots, \textbf{k}_n\right) \exp\left[f(k)D^2(z)\right] \, ,
\label{gamman}
\end{equation}
where the function $f$ depends only on the linear power spectrum today:
\begin{multline} \label{eq:fk}
f\left(k\right)=\int \frac{d^3q}{(2\pi)^3}\frac{P_{\text{lin}}\left(q,z=0\right)}{504k^3q^5} \left[ 6k^7q-79k^5q^3 +50q^5k^3 \right.  \\
\left.- 21kq^7 + \frac{3}{4}\left(k^2-q^2\right)^3\left(2k^2+7q^2\right)\log \frac{|k-q|^2}{|k+q|^2} \right] \, .
\end{multline}

The bispectrum contributions can be calculated from their SPT counterparts \cite{Bernardeau2012}, and the result up to two loops is
\begin{align} \label{eq:RPT_B}
B^{\text{MPTbreeze}}\left(k_1,k_2,k_3,z\right)=& \left[B_{\text{tree}}^{\mathrm{SPT}}  +(B_{222}+B_{321}^I)
+ (B_{134}^{II}+B_{224}^{II}+B_{233}^{II})\right]\left(k_1,k_2,k_3,z\right) \nonumber \\
&\times\exp\left[\left(f(k_1)+f(k_2)+f(k_3)\right)D^2(z)\right] \, .
\end{align}

The two-loop bispectrum has first been calculated numerically in Ref. \cite{Lazanu2016}. We extend these results to three loops using the bispectrum generating function of Ref. \cite{Bernardeau2008}. In this case there are four terms,

\begin{align}
\label{b3014}
&B_{014}^{\text{MPT}}(k_1,k_2,k_3)= 120 \int_{\bq,\br, \bs} F_5^{(s)}(-\bk_1,\bq,\br,\bs,-\bk_3-\bq-\br-\bs)  \nonumber \\
&  \quad \times   F_4^{(s)}(-\bq,-\br,-\bs,\bk_3+\bq+\br+\bs)  \Pl(q)\Pl(r)\Pl(s)\Pl(k_1)\Pl(|\bk_3+\bq+\br+\bs|) + 5 \text{ perms.} \, ,\\
\label{b3023}
&B_{023}^{\text{MPT}}(k_1,k_2,k_3)= 120 \int_{\bq,\br, \bs} F_5^{(s)}(\bq,\br,-\bk_3-\bq-\br,\bs,-\bk_2-\bs)F_2^{(s)}(-\bs,\bk_2+\bs)  \nonumber \\
& \quad  \times   F_3^{(s)}(-\bq,-\br,\bk_3+\bq+\br) \Pl(q) \Pl(r)\Pl(s)\Pl(|\bk_3+\bq+\br|)\Pl(|\bk_2+\bs|) + 5 \text{ perms.} \, ,\\
\label{b3113}
&B_{113}^{\text{MPT}}(k_1,k_2,k_3)= 192 \int_{\bq,\br, \bs} F_4^{(s)}(\bq,\br,\bs,\bk_1-\bq-\br-\bs) \nonumber \\
&   \quad \times F_2^{(s)}(-\bk_1+\bq+\br+\bs,-\bk_3-\bq-\br-\bs) F_4^{(s)}(\bk_3+\bq+\br+\bs,-\bq,-\br,-\bs) \nonumber \\
&   \quad \times    \Pl(q) \Pl(r)\Pl(s)\Pl(|\bk_1-\bq-\br-\bs|)\Pl(|\bk_3+\bq+\br+\bs|) + 2 \text{ perms.} \, ,\\
\label{b3122}
&B_{122}^{\text{MPT}}(k_1,k_2,k_3)= 216 \int_{\bq,\br, \bs} F_4^{(s)}(\bq,\br,\bs,\bk_1-\bq-\br-\bs) \nonumber \\
& \quad \times F_3^{(s)}(-\bs,-\bk_1+\bq+\br+\bs,-\bk_3-\bq-\br) F_3^{(s)}(\bk_3+\bq+\br,-\bq,-\br) \nonumber \\
&  \quad \times     \Pl(q)\Pl(r)\Pl(s)\Pl(|k_1-\bq-\br-\bs|)\Pl(|k_3+\bq+\br|) + 2 \text{ perms.}
\end{align}
\noindent
and hence the three-loop \textsc{MPTbreeze} bispectrum is given by
\begin{align}
\label{b3loopmpt}
B_{\text{3-loops}}^{\mathrm{MPT}}(k_1,k_2,k_3,z) = D^{10}(z) (B_{014}^{\text{MPT}} +B_{023}^{\text{MPT}}+B_{113}^{\text{MPT}}+B_{122}^{\text{MPT}}) (k_1,k_2,k_3) \nonumber \\
\times \exp[D^2(z)(f(k_1)+f(k_2)+f(k_3))]\, .
\end{align}

\section{IR-safe integrands}
From Eqs. (\ref{alpha}) and (\ref{beta}) it can be seen that the leading divergences of the kernel functions and of the integrands of the terms of Eq. (\ref{b2loop}) appear when their arguments are zero. However, by summing over all the terms prior to integration, these divergences cancel out due  to the Galilean  invariance of the equations of motion \cite{Jain1996, Scoccimarro1996}.
Therefore, in order to increase the precision of our computation, we aim to remove as many of the divergences involved prior to integration. This procedure involves taking advantage of the symmetries of the integrands and moving the leading divergences to a single point and it has been so far applied to the power spectrum \cite{Carrasco2014} and bispectrum \cite{Angulo2015,Lazanu2016}.
We therefore use the same approach in this work for the two-loop matter bispectrum. In Appendix \ref{sec:2lIR} we present the infrared (IR)-safe integrals for each of the 11 terms of the two-loop SPT bispectrum. To simplify the notation we denote the integrand of each term by its corresponding lowerscript letter and we only show the term shown explicitly, as the other permutations can be evaluated similarly. In each of the cases, we move all the leading divergences to $\bq=\br=0$. 

In Appendix \ref{sec:3lIR} we show a similar procedure for the three-loop \textsc{MPTbreeze} integrals, this time moving the divergences to $\bq=\br=\bs=0$.

\section{Results and comparison with simulations}

\subsection{Simulations}

In order to analyse the performance of each of the models considered, \textit{i.e.} SPT up to two loops and \textsc{MPTbreeze} up to three loops, we employ two sets of Gaussian simulations.
\begin{itemize}
 \item A set of simulations described in detail in Ref. \cite{Schmittfull2013}. They assume Gaussian 2LPT initial conditions and are evolved from $z=49$ until today using the \textsc{Gadget}-3 code \cite{Springel2001,Springel2005}. The simulations contain $512^3$ particles in a box size of $1600\, \text{Mpc}/h$, yielding a wavevector range of $[0.0039, 0.5]\, h/\text{Mpc}$. There are three realisations available for the simulations. The cosmology considered is a flat $\Lambda$CDM universe with the following WMAP7 \cite{Komatsu2011} parameters: baryon energy density $\Omega_bh^2=0.0226$, dark matter energy density $\Omega_ch^2=0.11$, cosmological constant energy density $\Omega_\Lambda=0.734$, dimensionless Hubble constant $h=0.71$, optical depth $\tau=0.088$, amplitude of primordial perturbations $\Delta^2_\mathcal{R}(k_0)=2.43\times 10^{-9}$ and scalar spectral index $n_s(k_0)=0.963$, where $k_0 = 0.002 \, h/  \mathrm{Mpc}$. The bispectrum was extracted from the simulations using the modal method, described in Refs. \cite{Fergusson2010,Fergusson2012,Regan2012} and around 100 modes have been used for the reconstruction. In this work we reconstruct the bispectrum from the modal functions. 

  \item Another set of simulations \cite{Desjacques2009,Sefusatti2010} with a box size of  $1600\, \text{Mpc}/h$ and a force resolution of 0.04 times the mean inter-particle distance, providing a wavevector range of $[0.012, 0.3]\, h/\text{Mpc}$. The initial distribution of particles is generated at $z=99$ using the Zel'dovich approximation and is evolved with the \textsc{Gadget} code. The cosmology is fixed to a flat $\Lambda$CDM universe with WMAP5 \cite{Komatsu2009} parameters: $h= 0.7$, $\Omega_m= 0.279$, $\Omega_b=0.0462$, $n_s= 0.96$ and a normalisation of the curvature perturbations $\Delta^2_R= 2.21 \times 10^{−9}$ (at $k = 0.02 \text{ Mpc}^{-1}$), yielding $\sigma_8 \approx 0.81$. Eight runs of the simulations have been performed. 
\end{itemize}

\subsection{Results and discussion}
\label{sec:res}

\begin{figure}[tbp]
\centering
\includegraphics[width=0.98\linewidth, trim={1.0cm 0 2.0cm 0},clip]{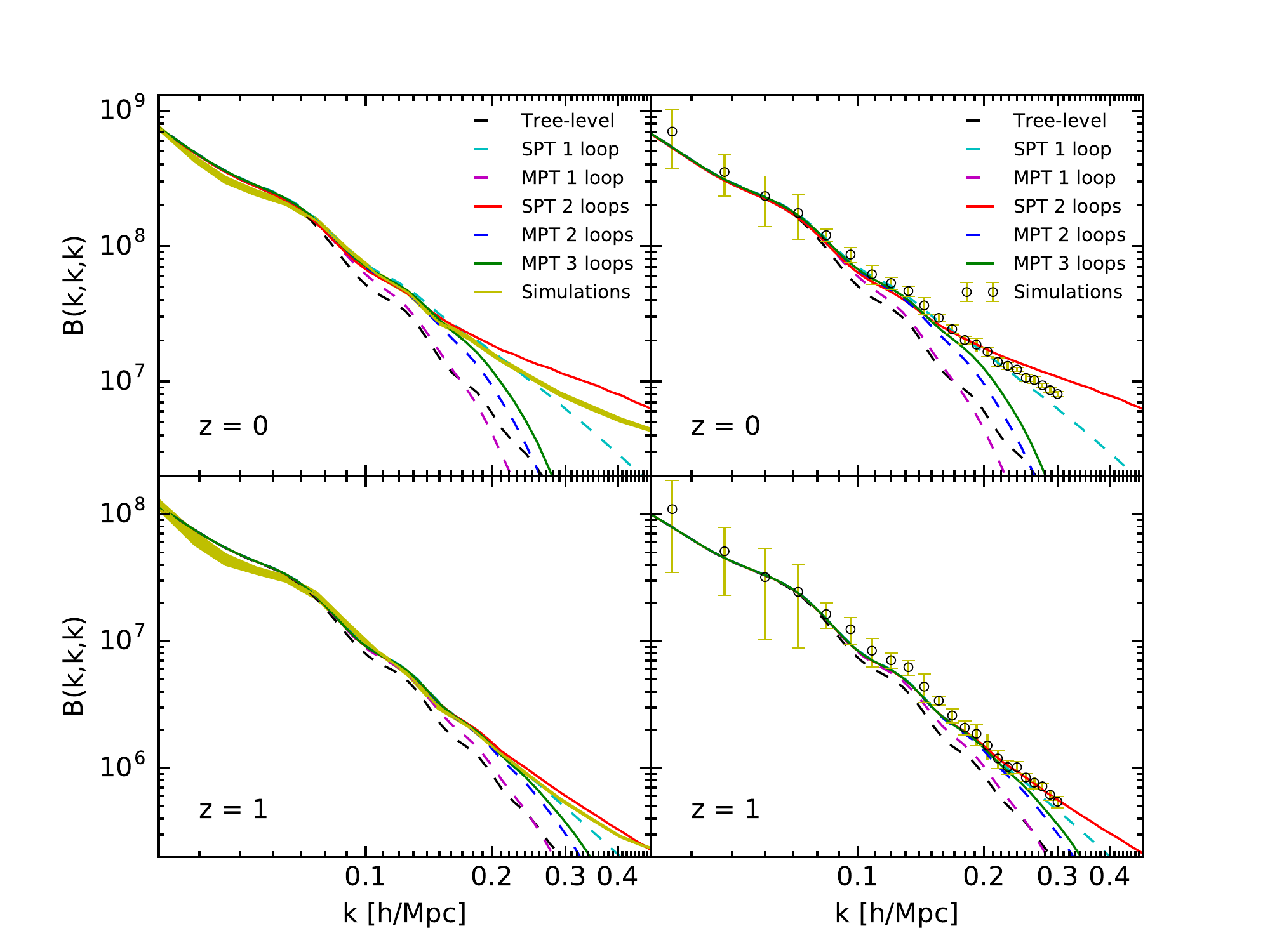} 
\caption{The matter bispectrum in the equilateral configuration at redshifts $z = 0$ (top) and $z = 1$ (bottom), showing SPT up to two loops, \textsc{MPTbreeze} up to three-loops and the $N$-body simulations of Ref. \cite{Schmittfull2013} (left) and of Ref. \cite{Sefusatti2010} (right).}
\label{fig:eq}
\end{figure}

In this Section, we compare the two-loop SPT results with their one-loop counterparts and with \textsc{MPTbreeze} at one, two and three loops as well as with $N$-body simulations at redshifts 0 and 1.
We compute the bispectra numerically using the \textsc{Cuba} Monte-Carlo integrator \cite{Hahn2005} in three (one loop), six (two loops) and nine  dimensions (three loops) respectively. We calculate the bispectra for the following theories: linear (tree-level), one and two loops in SPT, one, two and three loops in \textsc{MPTbreeze}. We evaluate the power spectrum with \textsc{Camb} \cite{Lewis2000} with the same cosmological parameters as the simulations. In the plots we also show the bispectrum extracted from $N$-body simulations (in yellow). In the case of simulations from Ref. \cite{Schmittfull2013}, we plot a band representing the uncertainties arising from the realisations of the simulations because they are reconstructed from basis functions. For the simulations of Ref. \cite{Sefusatti2010}, simulations are shown as points with error bars where the measurements have been made. Our conclusions regarding the validity of the various perturbation theories are in good agreement using the two simulation sets. We plot three configurations, equilateral, flattened and squeezed, corresponding to the cases usually investigated. The dashed lines represent results already available in the literature \cite{Bernardeau2002,Sefusatti2009,Lazanu2016,Lazanu2017}, while the continuous lines show the new results in this work, the bispectra at two loops in SPT  and at three loops in \textsc{MPTbreeze}. 

Compared to SPT, the \textsc{MPTbreeze} method is significantly less computationally expensive, due to the fewer number of terms required and to the lower order of the kernels: at two loops, SPT requires the sixth order kernel, compared to only the fourth in \textsc{MPTbreeze}, while at three loops the eighth order kernel appears in SPT, but only the fifth in \textsc{MPTbreeze}.

In the equilateral configuration (Fig. \ref{fig:eq}), the 2-loop SPT bispectrum becomes more accurate than its one loop counterpart, matching the simulations for $k \lesssim 0.14 \, h/\text{Mpc}$ at $z = 0$, curing the slight excess of power from the 1-loop SPT. For \textsc{MPTbreeze}, we inspect the bispectra at one, two and three loops and we observe that going from one to two loops the scale where the model is decaying is increased more compared to going from two to three loops. This behaviour was expected, since the expansion is convergent and therefore the additional contributions should indeed decrease with the loop order. \textsc{MPTbreeze} at one loop is accurate up $k \lesssim 0.08 \, h/\text{Mpc}$ at $z = 0$, while at two and three loops this is increased up to $k \lesssim 0.14 \, h/\text{Mpc}$ and $k \lesssim 0.17 \, h/\text{Mpc}$, scales similar to the two-loop SPT scenario. At $z = 1$, SPT provides similar predictions at one and two loops, up to $k \sim 0.2 \, h/\text{Mpc}$, and for smaller scales the two-loop SPT yields a light excess in signal.  \textsc{MPTbreeze} is accurate up to $k \sim 0.11 \, h/\text{Mpc}$, $k \sim 0.19 \, h/\text{Mpc}$ and $k \sim 0.24 \, h/\text{Mpc}$ at one, two and three loops respectively.

\begin{figure}[tbp]
\centering
\includegraphics[width=0.98\linewidth, trim={1.0cm 0 2.0cm 0},clip]{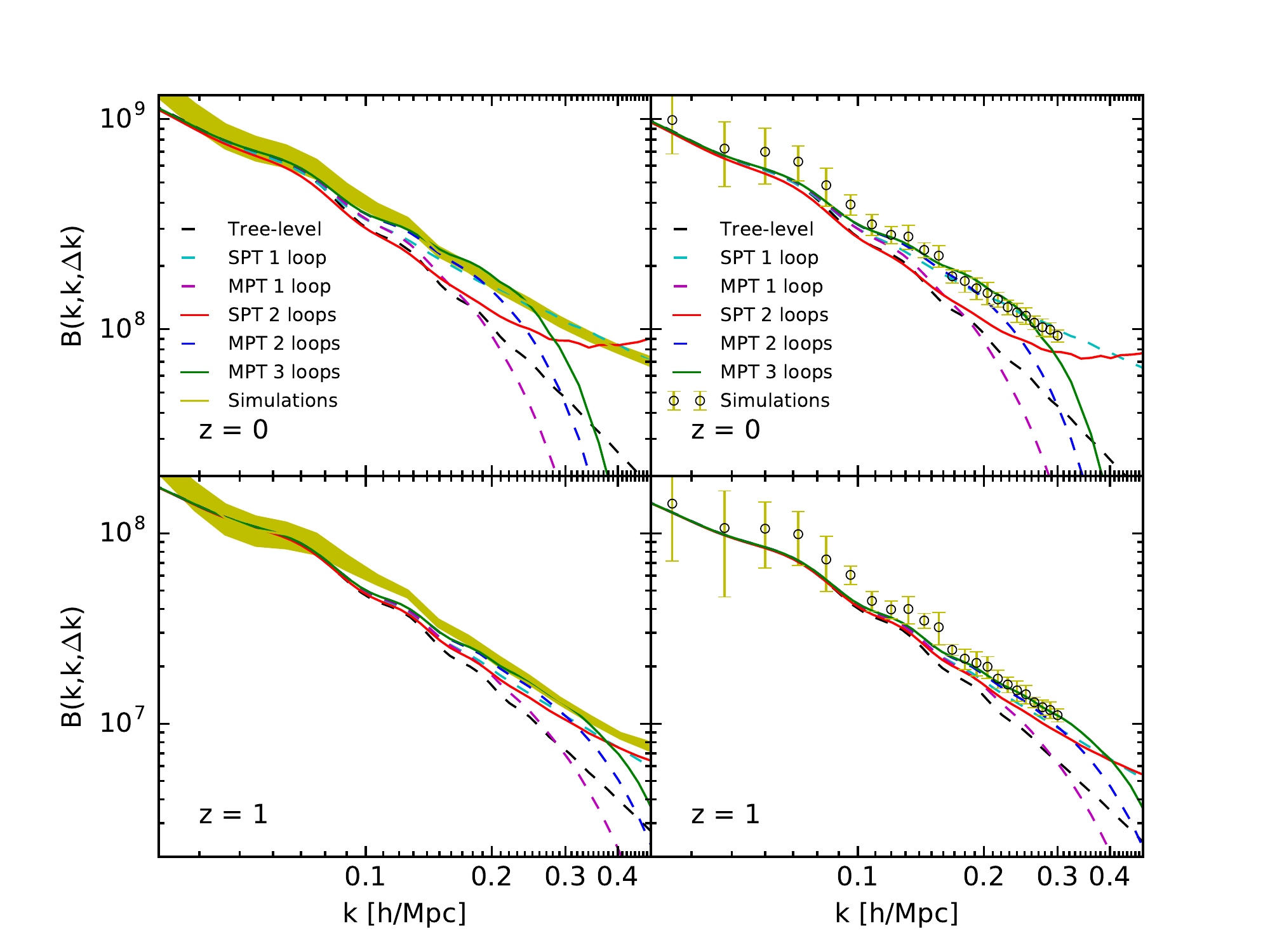} 
\caption{The matter bispectrum in the squeezed configuration (fixed small side $\Delta k \equiv 0.012 \, h/\text{Mpc}$) at redshifts $z = 0$ (top) and $z = 1$ (bottom), showing SPT up to two loops, \textsc{MPTbreeze} up to three-loops and the $N$-body simulations of Ref. \cite{Schmittfull2013} (left) and of Ref. \cite{Sefusatti2010} (right).}
\label{fig:sq}
\end{figure}

For the squeezed configuration (Fig. \ref{fig:sq}), we choose to consider triangles with sides $(k,k,\Delta k)$, with $\Delta k \equiv 0.012 \,h/\text{Mpc}$ and we plot the bispectra as functions of $k$. The simulations of Ref. \cite{Schmittfull2013} have larger error bars than in the equilateral case because one triangle side is always short, where error bars are larger. In this configuration, the one-loop SPT bispectrum has a good agreement with the simulations, while at two-loop it misses signal, having a lower magnitude than its one-loop counterpart for $k \lesssim 0.4 \, h/\text{Mpc}$ and it diverges for smaller scales at $z=0$. In the case of \textsc{MPTbreeze}, the model is providing a good description of the simulations for $k \lesssim 0.10 \, h/\text{Mpc}$, $k \lesssim 0.19 \, h/\text{Mpc}$ and $k \lesssim 0.25 \, h/\text{Mpc}$ at one, two and three-loops respectively. At $z = 1$, the two-loop SPT is still mildly less accurate than its one loop counterpart. The one-loop \textsc{MPTbreeze} matches the simulations for $k \lesssim 0.19 \, h/\text{Mpc}$, while the two and three-loops improve this to $k \lesssim 0.28 \, h/\text{Mpc}$ and $k \lesssim 0.33 \, h/\text{Mpc}$.

\begin{figure}[tbp]
\centering
\includegraphics[width=0.98\linewidth, trim={1.0cm 0 2.0cm 0},clip]{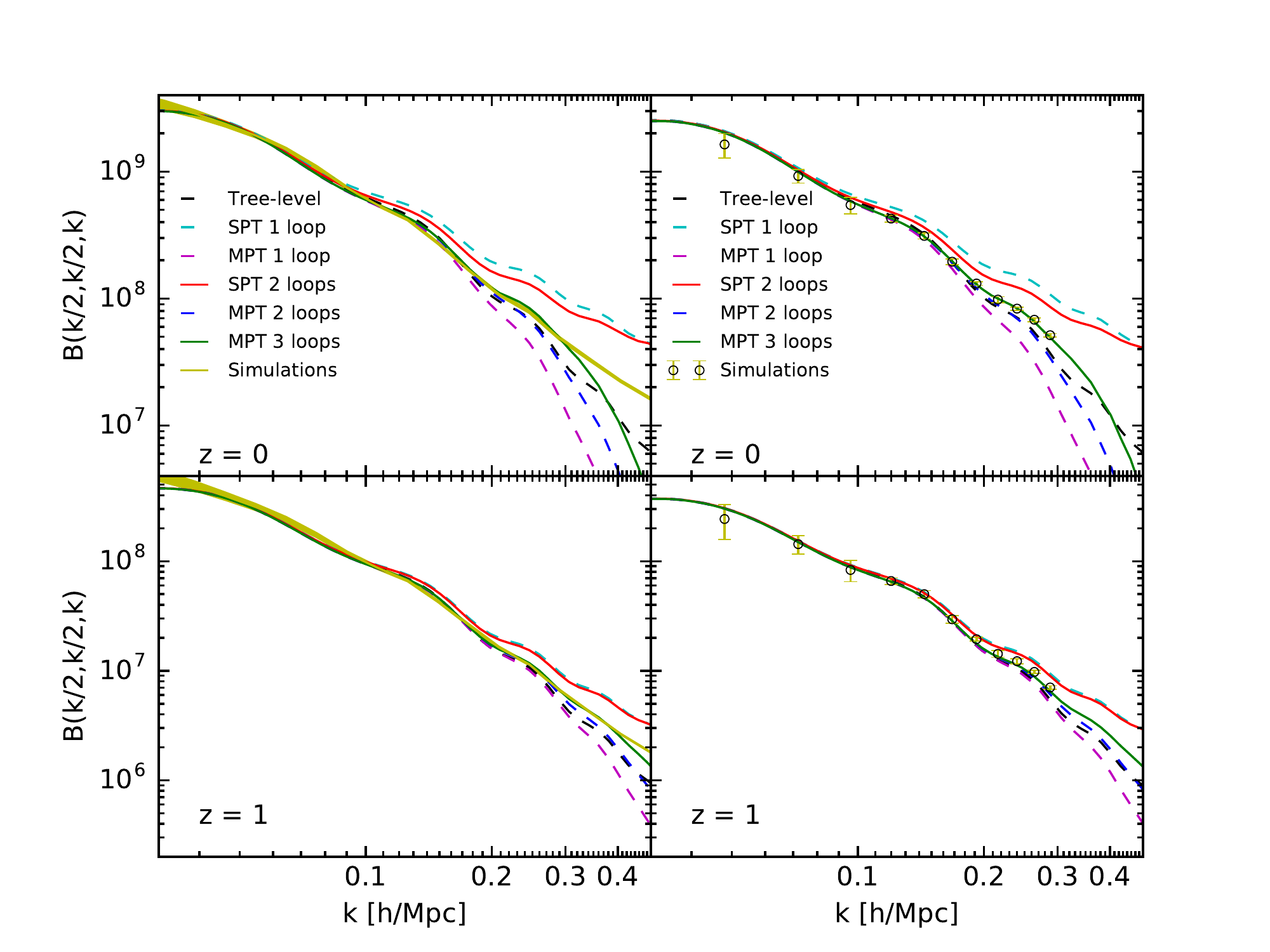} 
\caption{The matter bispectrum in the flattened configuration at redshifts $z = 0$ (top) and $z = 1$ (bottom), showing SPT up to two loops, \textsc{MPTbreeze} up to three-loops and the $N$-body simulations of Ref. \cite{Schmittfull2013} (left) and of Ref. \cite{Sefusatti2010} (right).}
\label{fig:fl}
\end{figure}

To be consistent with the plotting of bispectra in terms of the largest triangle side, we represent the flattened configuration (Fig. \ref{fig:fl}) as triangles of sides $(0.5k,0.5k,k)$. In this configuration, at $z = 0$ SPT presents an excess of signal at one loop on scales smaller than $k \sim 0.10 \,  h/\text{Mpc}$. At two loops, the excess is slightly diminished, but its amplitude is still higher than the predictions from $N$-body simulations for $k \gtrsim 0.09\,h/\text{Mpc}$. For \textsc{MPTbreeze} at one, two and three loops, we observe accurate fits for maximum scales of $k \lesssim 0.15 \, h/\text{Mpc}$, $k \lesssim 0.16 \, h/\text{Mpc}$ and $k \lesssim 0.28 \, h/\text{Mpc}$ respectively. At $z = 1$, the additional contribution to the SPT matter bispectrum coming from the terms at two loops is small, with bispectra both at one and two loops presenting excessive signals for $k \gtrsim 0.10 \, h/\text{Mpc}$. \textsc{MPTbreeze} is accurate for scales $k \lesssim 0.16 \, h/\text{Mpc}$ at one loop,  $k \lesssim 0.27 \, h/\text{Mpc}$ at two loops and $k \lesssim 0.36 \, h/\text{Mpc}$ at three loops.

Therefore, we see that \textsc{MPTbreeze} at three loops can accurately predict the matter bispectrum for $\max(k_1,k_2,k_3) \lesssim 0.17 \, h/\text{Mpc}$ at $z = 0$ in all configurations. There is only a small gain in the accuracy of the predictions by going from two to three loops ($\max(k_1,k_2,k_3) \lesssim 0.14 \, h/\text{Mpc}$ at two loops), while going from one to two loops yields a much more significant improvement -- the one-loop prediction can only provide a good modelling up to  $\max(k_1,k_2,k_3) \lesssim 0.08 \, h/\text{Mpc}$ at $z = 0$. At $z = 1$, the range of \textsc{MPTbreeze} increases up to $\max(k_1,k_2,k_3) \lesssim 0.24 \, h/\text{Mpc}$.

\section{Conclusion}
The bispectrum of LSS has so far been exploited much less than the power spectrum, although this quantity could be used for uncovering new physics and for breaking degeneracies between cosmological parameters.

In this work we have determined the dark matter bispectrum at two loops in SPT and at three loops in the renormalised \textsc{MPTbreeze} technique and we have compared them to two sets of $N$-body simulations. In the process we have removed divergences from the integrands, thus significantly reducing the risk of obtaining inaccurate results. The comparison to numerical simulations has shown that the two-loop SPT bispectrum does  increase the match to simulations with respect to the one loop case,  but requiring a much higher computational cost. The \textsc{MPTbreeze} approach is providing a convergent expansion and is therefore expected to provide a better match to simulations with the loop order. For scales larger than $k_1,k_2,k_3 \sim 0.17 \, h/\text{Mpc}$ at $z  = 0$, the three-loop bispectrum yields provides an adequate description of the simulations. The two-loop case can be used for  $k_1,k_2,k_3 \lesssim 0.14 \, h/\text{Mpc}$ at $z = 0$. At $z=1$, the three-loop \textsc{MPTbreeze} bispectrum is accurate up to $k_1,k_2,k_3 \lesssim 0.24 \, h/\text{Mpc}$ at $z = 1$, a gain from $0.19\, h/\text{Mpc}$ at two loops.

The two-loop SPT bispectrum can be used to calculate the two-loop Effective Field Theory of LSS bispectrum, which has a potential to go much further into the nonlinear regime. This would nevertheless require calculating the relevant counterterms and matching their amplitudes to $N$-body simulations. This work can also be generalised to include the effects of non-Gaussian initial conditions in both SPT and Effective Field Theory.

\appendix

\section{Two-loop SPT IR-safe integrands}
\label{sec:2lIR}
In this section we show how to remove the divergences in the two-loop SPT integrals.

\subsection*{$B_{116}$, $B_{125}^{I}$, $B_{134}^{III}$, $B_{233}^{III}$} This terms are already IR-safe, and no processing is required.

\subsection*{$B_{125}^{II}$} The leading divergences appear at $\bq=0$, $\bq=\bk_2$ and $\br=0$. The only divergence that needs removing is the one at $\bq=\bk_2$. Therefore, one can express the integral schematically as
    \begin{align}
     B_{125}^{II}=\int_{\bq,\br} b_{125}^{II} = 2 \int_{\bq,\br} b_{125}^{II} \Theta (|\bk_2-\bq|-q) \, ,
    \end{align}
    where $\Theta$ is the Heaviside function.

\subsection*{$B_{134}^{I}$} The leading divergences appear at $\bq=0$, $\bq=-\bk_3$ and $\br=0$. Then
     \begin{align}
     B_{134}^{I}&=\int_{\bq,\br} b_{134}^{I} 
     = 2 \int_{\bq,\br} b_{134}^{I} \Theta (|\bk_3+\bq|-q) \, .
    \end{align}
\subsection*{$B_{134}^{II}$, $B_{224}^{II}$, $B_{233}^{II}$} This terms have been investigated in Ref. \cite{Lazanu2016} (Eqs. E9, E10 and E11).

\subsection*{$B_{224}^{I}$} \label{sec:224i} The leading divergences appear at $\bq=0$, $\bq=\bk_1$, $\bq=-\bk_2$ and $\br=0$. We aim to remove the divergences at $\textbf{q}=\bk_1$ and $\bq=-\bk_2$ by moving them to $\bq=0$. We thus split the $\bq$ integration domain into four parts, by multiplying the integral with a products of two Heaviside functions, as follows:
  
\begin{itemize}
 \item $\Theta(|\bk_1-\bq|-q)\Theta(|\bk_2+\bq|-q)$. 
In this case, the divergences at $\textbf{q}=\bk_1$ and $\bq=-\bk_2$ are removed since the Heaviside functions evaluate to 0 at these points.
  \item $\Theta(|\bk_1-\bq|-q)\Theta(q-|\bk_2 +\bq|)$,
  Here, the divergence at $\textbf{q}=\bk_1$ is eliminated, but not the one at $\bq=-\bk_2$. Hence, we perform the change of variable $\tilde{\bq}=-\bk_2-\bq$ and the integrand becomes
     \begin{align}
      &b_{224}^{I}=48 F_2^{(s)}(-\bk_2-\tilde{\bq},-\bk_3+\tilde{\bq})F_2^{(s)}(-\tilde{\bq},\bk_2+\tilde{\bq}) \nonumber \\
      &\times F_4^{(s)}(\bk_3-\tilde{\bq},\tilde{\bq},\br,-\br) \Pl(|\tilde{\bq}-\bk_3|) \Pl(\tilde{\bq}) \Pl(|\bk_2+\tilde{\bq}|) \nonumber \\
&\times \Pl(r) \Theta(|-\bk_3+\tilde{\bq}|-|\bk_2+\tilde{\bq}|) \Theta(|\bk_2+\tilde{\bq}|-\tilde{q}) \, .
     \end{align}
      Divergences are now at $\tilde{\bq}=0$, $\tilde{\bq}=\bk_3$, $\tilde{\bq}=-\bk_2$ and $\br=0$, but the only remaining ones remain at $\tilde{\bq}=\br=0$ since the Heaviside functions evaluate to zero for the other two points.
  \item $\Theta(q-|\bk_1-\bq|)\Theta(|\bk_2+\bq|-q)$.
  The divergence at $\textbf{q}=-\bk_2$ is eliminated, but not the one at $\bq=\bk_1$. Hence, we perform the change of variable $\tilde{\bq}=\bk_1-\bq$ and the integrand becomes
     \begin{align}
      &b_{224}^{I}=48 F_2^{(s)}(\bk_1-\tilde{\bq},\tilde{\bq})F_2^{(s)}(-\bk_3-\tilde{\bq},-\bk_1+\tilde{\bq}) \nonumber \\
      &\times F_4^{(s)}(-\tilde{\bq},\bk_3+\tilde{\bq},\br,-\br) \Pl(|\tilde{\bq}|) \Pl(|\bk_3+\tilde{\bq}|) \Pl(|\bk_1-\tilde{\bq}|) \nonumber \\
&\times \Pl(r) \Theta(|\bk_1-\tilde{q}||) \Theta(|\bk_3+\tilde{\bq}|-|\bk_1-\tilde{\bq}|) \, .
     \end{align}
      Divergences are now at $\tilde{\bq}=0$, $\tilde{\bq}=\bk_1$, $\tilde{\bq}=-\bk_3$ and $\br=0$, but the only remaining ones remain at $\tilde{\bq}=\br=0$ since the Heaviside functions evaluate to zero for the other two points.
  \item $\Theta(q-|\bk_1-\bq|)\Theta(q-|\bk_2+\bq|)$. 
      We perform the same change of variable as above, but in this case the divergences at $\tilde{\bq}=\bk_1$ and $\tilde{\bq}=-\bk_3$ are only eliminated when $k_3>k_2$. This is not satisfactory and hence we split the integration domain again with two Heaviside functions, $\Theta(|\bk_3+\tilde{\bq}|-\tilde{q})$ and $\Theta(\tilde{q}-|\bk_3+\tilde{\bq}|)$. In the former case, divergences at points different from zero are removed and in the latter one a change of variable $\tilde{\tilde{\bq}}=\bk_3+\tilde{\bq}$ solves the problem.
\end{itemize}

Therefore, the final expression for the first permutation can be expressed as
\begin{align}
B_{224}^I(&k_1,k_2,k_3)=\nonumber \\
=48 & \int_{\bq,\br} F_2^{(s)}(\bq,\bk_1-\bq)F_2^{(s)}(-\bq,\bk_2+\bq) F_4^{(s)}(\bq-\bk_1,-\bk_2-\bq,\br,-\br) \nonumber \\
&\times \Pl(|\bk_1-\bq|) \Pl(|\bk_2+\bq|)\Pl(q)\Pl(r) \Theta(|\bk_1-\bq|-q)\Theta(q-|\bk_2 +\bq|) \nonumber  \\
+ 48 & \int_{\tilde{\bq},\br}F_2^{(s)}(\bk_1-\tilde{\bq},\tilde{\bq})F_2^{(s)}(-\bk_3-\tilde{\bq},-\bk_1+\tilde{\bq}) F_4^{(s)}(-\tilde{\bq},\bk_3+\tilde{\bq},\br,-\br) \nonumber \\
&      \times  \Pl(\tilde{q}) \Pl(|\bk_3+\tilde{\bq}|) \Pl(|\bk_1-\tilde{\bq}|) \Pl(r) \Theta(|\bk_1-\tilde{q}||) \Theta(|\bk_3+\tilde{\bq}|-|\bk_1-\tilde{\bq}|) \nonumber \\
+48 &\int_{\tilde{\bq},\br} F_2^{(s)}(\bk_1-\tilde{\bq},\tilde{\bq})F_2^{(s)}(-\bk_3-\tilde{\bq},-\bk_1+\tilde{\bq})F_4^{(s)}(-\tilde{\bq},\bk_3+\tilde{\bq},\br,-\br) \nonumber \\
&      \times  \Pl(\tilde{q}) \Pl(|\bk_3+\tilde{\bq}|) \Pl(|\bk_1-\tilde{\bq}|)  \Pl(r) \Theta(|\bk_1-\tilde{q}||) \Theta(|\bk_3+\tilde{\bq}|-|\bk_1-\tilde{\bq}|) \nonumber \\
+48 &\int_{\tilde{\bq},\br} F_2^{(s)}(\bk_1-\tilde{\bq},\tilde{\bq})F_2^{(s)}(-\bk_3-\tilde{\bq},-\bk_1+\tilde{\bq})F_4^{(s)}(-\tilde{\bq},\bk_3+\tilde{\bq},\br,-\br) \Pl(\tilde{q}) \nonumber \\
&      \times   \Pl(|\bk_3+\tilde{\bq}|) \Pl(|\bk_1-\tilde{\bq}|)  \Pl(r) \Theta(|\bk_1-\tilde{q}||) \Theta(|\bk_3+\tilde{\bq}|-|\bk_1-\tilde{\bq}|)  \Theta(|\bk_3+\tilde{\bq}|-\tilde{q})  \nonumber \\
+48 &\int_{\tilde{\tilde{\bq}},\br} F_2^{(s)}(-\bk_2-\tilde{\tilde{\bq}},\tilde{\tilde{\bq}}-\bk_3)F_2^{(s)}(\bk_2+\tilde{\tilde{\bq}},-\tilde{\tilde{\bq}})F_4^{(s)}(-\tilde{\tilde{\bq}}+\bk_3,\tilde{\tilde{\bq}},\br,-\br) \Pl(|\tilde{\tilde{\bq}}-\bk_3|)  \Pl(\tilde{\tilde{q}}) \nonumber \\
&      \times   \Pl(|\bk_2+\tilde{\tilde{q}}|)  \Pl(r) \Theta(|\bk_2+\tilde{\tilde{q}}|-|\tilde{\tilde{\bq}}-\bk_3||) \Theta(|\bk_2+\tilde{\tilde{\bq}}|-\tilde{\tilde{q}}) \Theta(|\tilde{\tilde{q}}-\bk_3|-\tilde{\tilde{q}}) \, .
\end{align}

 \subsection*{$B_{233}^{I}$} The leading divergences appear at $\textbf{q}=0$, $\textbf{q}=\bk_1$ and $\br=0$.  Then,
     \begin{align}
     B_{233}^{I}&=\int_{\bq,\br} b_{233}^{I} = 2 \int_{\bq,\br} b_{233}^{I} \Theta (|\bk_1-\bq|-q) \, .
    \end{align}
To eliminate subleading divergences, we employ a similar technique to Ref.  to symmetrise over angles, to make the expressions symmetric in $\bq \leftrightarrow \br$ and then to restrict the integration domain to $q>r$.

\section{Three-loop MPTbreeze IR-safe integrands}
\label{sec:3lIR}
In this Appendix we present a short summary of how to remove the leading divergences appearing in the four terms of the three-loop \textsc{MPTbreeze} bispectrum, using similar techniques to the ones used for SPT and for the two-loop \textsc{MPTbreeze} in Ref. \cite{Lazanu2016}.

\subsection*{$B_{014}^{\text{MPT}}$}
\label{MPT31}
We consider the integrand $b_{014}$ of the first permutation of $B_{014}^{\text{MPT}}$. The leading divergences of this term are at $\bq=0$, $\br=0$, $\bs=0$ and $\bq+\br+\bs=-\bk_3$. We  introduce an additional integration variable, $\bt=-\bq-\br-\bs-\bk_3$ and hence the integral becomes
\begin{align}
&B_{014}^{\text{MPT}}(k_1,k_2,k_3)_{\text{perm 1}}= 120 \int_{\bq,\br, \bs, \bt} F_5^{(s)}(-\bk_1,\bq,\br,\bs,\bt) F_4^{(s)}(-\bq,-\br,-\bs,-\bt)  \nonumber \\
&  \quad \times    \Pl(k_1)\Pl(q)\Pl(r)\Pl(s)\Pl(t) \delta_D(\bk_3+\bq+\br+\bs+\bt) \, .
\end{align}
This expression is now completely symmetrical under the pairwise exchange of $\bq$, $\br$, $\bs$, $\bt$ and hence we can consider an ordering $t>s>r>q$. As there are 24 permutations of the four variables, we can rewrite the integral assuming this ordering and remove the $\bt$ integration
\begin{align}
&B_{014}^{\text{MPT}}(k_1,k_2,k_3)= 120 \times 24 \int_{\bq,\br, \bs} F_5^{(s)}(-\bk_1,\bq,\br,\bs,-\bk_3-\bq-\br-\bs)  \nonumber \\
&  \quad \times   F_4^{(s)}(-\bq,-\br,-\bs,\bk_3+\bq+\br+\bs)  \Pl(q)\Pl(r)\Pl(s)\Pl(k_1)\Pl(|\bk_3+\bq+\br+\bs|) \nonumber \\
&  \quad \times \Theta(|\bk_3+\bq+\br+\bs|-s)\Theta(s-r)\Theta(r-q) + 5 \text{ perms.}  \, .
\end{align}

\subsection*{$B_{023}^{\text{MPT}}$}
\label{MPT32}
The leading divergences are at $\bq=0$, $\br=0$, $\bs=0$, $\bq+\br=-\bk_3$ and $\bs=-\bk_2$. As the integrand is symmetrical under the exchange $\bs \leftrightarrow -\bk_2-\bs$, we can remove the divergence at $\bs=-\bk_2$ by multiplying the expression by $2 \Theta(|\bs+\bk_2|-s)$. Similarly, the other divergence is removed by $2 \Theta(|\bq+\br+\bk_3|-|\bq+\br|) \times 2 \Theta(r-q)$, so that the term becomes

\begin{align}
&B_{023}^{\text{MPT}}(k_1,k_2,k_3)= 120 \times 8 \int_{\bq,\br, \bs} F_5^{(s)}(\bq,\br,-\bk_3-\bq-\br,\bs,-\bk_2-\bs)F_2^{(s)}(-\bs,\bk_2+\bs)  \nonumber \\
& \quad  \times   F_3^{(s)}(-\bq,-\br,\bk_3+\bq+\br) \Pl(q) \Pl(r)\Pl(s)\Pl(|\bk_3+\bq+\br|)\Pl(|\bk_2+\bs|) \nonumber \\
& \quad \times \Theta(|\bq+\br+\bk_3|-|\bq+\br|)  \Theta(r-q) \Theta(|\bs+\bk_2|-s) + 5 \text{ perms.} \, .
\end{align}

\subsection*{$B_{113}^{\text{MPT}}$}
In this case the leading divergences are at $\bq=0$, $\br=0$, $\bs=0$, $\bq+\br+\bs=\bk_1$ and $\bq+\br+\bs=-\bk_3$.  For the latter two divergences we introduce two additional integration variables, such that the integral becomes

\begin{align}
&B_{113}^{\text{MPT}}(k_1,k_2,k_3)_{\text{perm 1}}= 192 \int_{\bq,\br, \bs, \bt, \bu} F_4^{(s)}(\bq,\br,\bs,\bt) F_2^{(s)}(-\bt,-\bu) F_4^{(s)}(\bu,-\bq,-\br,-\bs) \nonumber \\
&    \, \times  \delta_D(\bk_1-\bq-\br-\bs-\bt) \delta_D(\bu- \bk_3-\bq-\br-\bs) 
   \Pl(q) \Pl(r)\Pl(s)\Pl(t)\Pl(u) \, .
\end{align}

The divergences appear when the arguments of the above kernels are 0. Because the integrand is symmetric under the exchange of $\bq$, $\br$ and $\bs$ as well as $\bt \leftrightarrow -\bu$,  we can consider an ordering of the magnitudes of the three wavevectors, $s \ge q \ge r$ and $t \ge u$ and multiply the integrand by 12 together with appropriate Heaviside functions. We note that there are 10 possible orderings of the five variables considering the previous constraints. Denoting

\begin{align}
 f(\bq,\br,\bs,\bt,\bu)&= 192 \times 12 \int_{\bq,\br, \bs, \bt, \bu} F_4^{(s)}(\bq,\br,\bs,\bt) F_2^{(s)}(-\bt,-\bu) \nonumber \\
 & \quad \times F_4^{(s)}(\bu,-\bq,-\br,-\bs) \Theta(s-q) \Theta(q-r) \Theta(t-u) \, ,
\end{align}
the bispectrum $B_{113}^{\text{MPT}}$ can be expressed in an IR-safe way as 

\begin{align}
&B_{113}^{\text{MPT}}(k_1,k_2,k_3)_{\text{perm 1}} \nonumber \\
&= \int_{\bq,\br, \bs} f(\bq,\br,\bs,\bk_1-\bq-\br-\bs,\bk_3+\bq+\br+\bs) \Theta(|\bk_1-\bq-\br-\bs|-|\bk_3+\bq+\br+\bs|) \nonumber \\
& \qquad \times \Theta(|\bk_3+\bq+\br+\bs|-s)\Theta(s-r)\Theta(r-q) \nonumber \\
&+ \int_{\bq,\br, \bu} f(\bq,\br,\bu-\bk_3-\bq-\br,-\bk_2-\bu,\bu) \Theta(|-\bk_2-\bu|-|\bu-\bk_3-\bq-\br|) \nonumber \\
& \qquad \times \Theta(|\bu-\bk_3-\bq-\br|-u)\Theta(u-r)\Theta(r-q) \nonumber \\
&+ \int_{\bq,\bs, \bu} f(\bq,\bu-\bk_3-\bq-\bs,\bs,-\bk_2-\bu,\bu) \Theta(|-\bk_2-\bu|-s) \nonumber \\
& \qquad \times \Theta(s-|\bu-\bk_3-\bq-\bs|)\Theta(|\bu-\bk_3-\bq-\bs|-u)\Theta(u-q)  \nonumber \\
&+ \int_{\bq,\bs, \bu} f(\bq,\bu-\bk_3-\bq-\bs,\bs,-\bk_2-\bu,\bu) \Theta(|-\bk_2-\bu|-s) \nonumber \\
& \qquad \times \Theta(s-|\bu-\bk_3-\bq-\bs|)\Theta(|\bu-\bk_3-\bq-\bs|-q)\Theta(q-u) \nonumber \\
&+ \int_{\bq,\br, \bt} f(\bq,\br,\bk_1-\bq-\br-\bt,\bt,\bu) \Theta(|\bk_1-\bq-\br-\bt|-t) \nonumber \\
& \qquad \times \Theta(t-|-\bk_2-\bt|)\Theta(|-\bk_2-\bt|-r)\Theta(r-q) \nonumber \\
&+ \int_{\bq,\br, \bs} f(\bq,\br,\bs,\bk_1-\bq-\br-\bs,\bu) \Theta(s-|\bk_1-\bq-\br-\bs|) \nonumber \\
& \qquad \times \Theta(|\bk_1-\bq-\br-\bs|-r)\Theta(r-|\bk_3+\bq+\br+\bs|)\Theta(|\bk_3+\bq+\br+\bs|-q) \nonumber \\
&+ \int_{\br,\bs, \bu} f(\bu-\bk_3-\br-\bs,\br,\bs,-\bk_2-\bu,\bu) \Theta(s-|-\bk_2-\bu|) \nonumber \\
& \qquad \times \Theta(|-\bk_2-\bu|-r)\Theta(r-|\bu-\bk_3-\br-\bs|)\Theta(|\bu-\bk_3-\br-\bs|-u) \nonumber \\
&+ \int_{\bq,\br, \bu} f(\bq,\br,\bu-\bk_3-\bq-\br,-\bk_2-\bu,\bu) \Theta(|\bu-\bk_3-\bq-\br|-r) \nonumber \\
& \qquad \times \Theta(r-|-\bk_2-\bu|)\Theta(|-\bk_2-\bu|-u)\Theta(u-q) \nonumber \\
&+ \int_{\bq,\br, \bu} f(\bq,\br,\bu-\bk_3-\bq-\br,-\bk_2-\bu,\bu) \Theta(|\bu-\bk_3-\bq-\br|-r) \nonumber \\
& \qquad \times \Theta(r-|-\bk_2-\bu|)\Theta(|-\bk_2-\bu|-q)\Theta(q-u) \nonumber \\
&+ \int_{\bq,\bs, \bu} f(\bq,\bu-\bk_3-\bq-\bs,\bs,-\bk_2-\bu,\bu) \Theta(s-|\bu-\bk_3-\bq-\bs|) \nonumber \\
& \qquad \times \Theta(|\bu-\bk_3-\bq-\bs|-q)\Theta(q-|-\bk_2-\bu|)\Theta(|-\bk_2-\bu|-u) \, .
\end{align}

We note that the integrand is symmetric under the exchange of $\bq$, $\br$ and $\bs$ and hence we can introduce an ordering of the magnitudes of the three wavevectors, together with a factor of 6. Therefore, we assume $s \ge r \ge q$. For the variables $t$ and $u$, we must consider all 20 possible orderings with respect to $q$, $r$ and $s$. 

\subsection*{$B_{122}^{\text{MPT}}$}
Leading divergences are at $\bq=0$, $\br=0$, $\bs=0$, $\bq+\br+\bs=\bk_1$ and $\bq+\br=-\bk_3$. The last two ones can be removed with Heaviside functions as in Sec. \ref{MPT32}, so that the integral becomes

\begin{align}
&B_{122}^{\text{MPT}}(k_1,k_2,k_3)= 216 \times 8\int_{\bq,\br, \bs} F_4^{(s)}(\bq,\br,\bs,\bk_1-\bq-\br-\bs) \nonumber \\
& \quad \times F_3^{(s)}(-\bs,-\bk_1+\bq+\br+\bs,-\bk_3-\bq-\br) F_3^{(s)}(\bk_3+\bq+\br,-\bq,-\br) \nonumber \\
&  \quad \times     \Pl(q)\Pl(r)\Pl(s)\Pl(|k_1-\bq-\br-\bs|)\Pl(|k_3+\bq+\br|) \nonumber \\
&  \quad \times  \Theta(|\bk_3+\bq+\br|-|\bq+\br|) \Theta(r-q) \Theta(|\bk_1-\bq-\br-\bs|-|\bq+\br+\bs|)  + 2 \text{ perms.} \,.
\end{align}

To eliminate subleading divergences, we symmetrise all the expressions in $\bq \leftrightarrow \br \leftrightarrow \bs$ and multiply the expression by $6 \Theta(s-r) \Theta(r-q)$ (unless this procedure has already been applied).

\acknowledgments
The authors are grateful to Marcel Schmittfull and Emiliano Sefusatti for providing the simulations used in the comparison. AL thanks Paul Shellard for illuminating discussions during the early stages of this work.
Cloud Area Padovana is acknowledged for the use of computing and storage facilities.

\bibliography{Biblio}{}

\end{document}